\documentclass[structabstract]{aa}  
\usepackage{graphicx}
\usepackage{txfonts}
\usepackage{natbib}
\usepackage{aalongtable}
\begin{document}
\title{Weak magnetic field, solid-envelope rotation, and wave-induced N-enrichment in the SPB star $\zeta$~Cassiopeiae\thanks{Based on observations obtained at the Telescope Bernard Lyot (USR5026) operated by the Observatoire Midi-Pyr\'en\'ees, Universit\'e de Toulouse (Paul Sabatier), Centre National de la Recherche Scientifique (CNRS) of France.}}

\titlerunning{$\zeta$~Cas: weak magnetic field, solid-envelope rotation and N-enrichment}

   \author{M. Briquet
          \inst{1,2}\fnmsep\thanks{F.R.S.-FNRS Postdoctoral Researcher, Belgium}
          \and
          C.  Neiner\inst{2}
          \and
          P. Petit\inst{3}
          \and 
          B. Leroy\inst{2}
          \and
          B. de Batz\inst{2}
          \and the MiMeS collaboration}

   \institute{
         Institut d'Astrophysique et de G\'eophysique, Universit\'e de Li\`ege, Quartier Agora, All\'ee du 6 ao\^ut 19C, B-4000 Li\`ege, Belgium\\
              \email{maryline.briquet@ulg.ac.be}
         \and
LESIA, Observatoire de Paris, PSL Research University, CNRS,
Sorbonne Universit\'es, UPMC Univ. Paris 06, Univ. Paris Diderot,\\
Sorbonne Paris Cit\'e, 5 place Jules Janssen, 92195 Meudon, France        
         \and
Institut de Recherche en Astrophysique et Plan\'etologie, Universit\'e de Toulouse UPS-OMP / CNRS, 14, avenue Edouard Belin, 31400 Toulouse, France
             }

   \date{Received ; accepted}

 
  \abstract
   {}
   {The main-sequence B-type star $\zeta$~Cassiopeiae is known as a N-rich star with a magnetic field discovered with the Musicos spectropolarimeter. We model the magnetic field of the star by means of 82 new spectropolarimetric observations of higher precision to investigate the field strength, topology, and effect.}
   {We gathered data with the Narval spectropolarimeter installed at T\'elescope Bernard Lyot (TBL, Pic du Midi, France) and applied the least-squares deconvolution technique to measure the circular polarisation of the light emitted from $\zeta$~Cas. We used a dipole oblique rotator model to determine the field configuration by fitting the longitudinal field measurements and by synthesizing the measured Stokes $V$ profiles. We also made use of the Zeeman-Doppler Imaging technique to map the stellar surface and to deduce the difference in rotation rate between the pole and equator.}
   {$\zeta$~Cas exhibits a polar field strength $B_{\rm pol}$ of 100-150 G, which is the weakest polar field observed so far in a massive main-sequence star. Surface differential rotation is ruled out by our observations and the field of $\zeta$~Cas is strong enough to enforce rigid internal rotation in the radiative zone according to theory. Thus, the star rotates as a solid body in the envelope. }
   {We therefore exclude rotationally-induced mixing as the cause of the surface N-enrichment. We discuss that the transport of chemicals from the core to the surface by internal gravity waves is the most plausible explanation for the nitrogen overabundance at the surface of $\zeta$~Cas.}


   \maketitle
%

\section{Introduction}
Among main-sequence B-type stars, it is well known that the Bp stars -- B-type stars showing abnormal abundances of certain chemical elements in their atmosphere -- are usually found to possess strong magnetic fields of several hundred Gauss to tens of kGauss. Weak magnetic fields of the order of a few hundreds Gauss have also been detected in several B-type targets that undergo stellar pulsations, i.e. in $\beta$~Cep stars and in slowly pulsating B (SPB) stars. The first detections of weak longitudinal magnetic fields (lower than 100 G) in the group of pulsating B stars have been achieved for the $\beta$~Cep stars $\beta$~Cephei \citep{henrichs2000} and V\,2052\,Oph \citep{neiner2003a}. Following these discoveries, searches for magnetic fields in $\beta$~Cep and SPB stars have been accomplished (e.g., \citealp{hubrig2006}; \citealp{hubrig2009}; \citealp{henrichs2009}; \citealp{silvester2009}; \citealp{alecian2011}; \citealp{briquet2013}; \citealp{neiner2014a}; \citealp{sodor2014}), definitely showing that magnetic fields are present among B-type pulsators. However, direct detections of a Zeeman signature in the Stokes $V$ measurements has only been obtained for $\sim$70 magnetic massive stars and they remain limited for pulsating B stars ($\sim$15 targets). Moreover, there are only a few detailed studies of the magnetic field configuration for these objects (e.g., \citealp{hubrig2011}; \citealp{henrichs2012}; \citealp{neiner2012}).

The B2\,IV-V star $\zeta$~Cas (HD~3360) was discovered to be magnetic with the Musicos spectropolarimeter at TBL (Pic du Midi, France) by \cite{neiner2003b}. Independent spectroscopic determinations of the fundamental parameters (\citealp{briquet2007}; \citealp{lefever2010}; \citealp{nieva2012}) place this  target in the common part of the $\beta$~Cep and SPB instability strips in the H-R diagram, making the star a candidate hybrid B-type pulsator. The fact that this object is a candidate pulsating target was also suggested from line profile studies by \cite{smith1980}, \cite{sadsaoud1994}, and \cite{neiner2003b}, successively. In the three studies, one non-radial pulsation g mode is observed but the suggested periodicity differs from one paper to the other. The more recent and more extended study by \cite{neiner2003b} leads to the detection of a non-radial pulsation mode with $\ell = 2 \pm 1$ at the frequency f = 0.64~d$^{-1}$ with a radial velocity amplitude of the order of 3 km~s$^{-1}$. In addition, a nitrogen overabundance is well known in $\zeta$~Cas (e.g., \citealp{gies1992}; \citealp{neiner2003b}; \citealp{nieva2012}), similarly to what is observed for other magnetic OB stars on the main sequence (\citealp{morel2008}; \citealp{martins2012}). 

In this paper, we present a new study of $\zeta$~Cas in order to better constrain its magnetic field, as was already achieved for V\,2052\,Oph \citep{neiner2012}, by making use of the high-efficiency and high-resolution Narval spectropolarimeter installed on the TBL 2-m telescope (Pic du Midi, France). The rotation period of the star ($P_{\rm rot}=5.370447$$\pm$$0.000078$~d) is known accurately thanks to a detailed study of time-resolved wind-sensitive UV resonance lines taken by the IUE satellite \citep{neiner2003b}. Our new spectropolarimetric measurements of $\zeta$~Cas were thus taken so that they cover well the various phases of rotation. These observations and the magnetic field measurements are described in Sect.\,\ref{obs} and Sect.\,\ref{magmeas}, respectively. A modelling of the magnetic field and of differential rotation is presented in Sect.\,\ref{magmod}, followed by a discussion in Sect.\,\ref{disc}. We end the paper with a summary of our conclusions in Sect.\,\ref{sum}.


\section{Observations}\label{obs}
Ninety six high-resolution (R = 65 000) spectropolarimetric Narval observations of $\zeta$~Cas were collected in 2007, 2009, 2011 and 2012 (see Table\,\ref{log}). Unfortunately, we had to discard 12 observations from our analysis (Nr.~46, 48, 49, 62, 63, 64, 65, 66, 67, 68, 69, 86). Indeed, anomalies in the behaviour of one of the rhombs of Narval have been detected during some of the observing nights in 2011 and 2012, due to the ageing of the coding disk of the rhomb. Because the rhomb was not in the right position, observations taken during these nights show a Stokes~$V$ signature that is either weaker or altogether absent. This technical problem has been solved by the TBL team in September 2012. Moreover, observations Nr.~1 and 2 were obtained in the linear polarisation mode (Stokes $Q$ and $U$) and are not used in our analysis which concerns circular polarisation (Stokes $V$). In total, we thus use 82 Stokes $V$ measurements in the present work. 

Each of the polarimetric measurements consists in a sequence of four subexposures with the same exposure time, between 120 and 420 seconds each, taken in different configurations of the polarimeter retarders. Usual bias, flat-fields, and ThAr calibrations were obtained at the beginning and at the end of each night. The data reduction was performed using {\sc libre-esprit} \citep{donati1997}, the dedicated reduction software available at TBL. From each set of four subexposures we derived one Stokes $I$ spectrum, as well as a Stokes $V$ spectrum and a null ($N$) polarisation spectrum by constructively and destructively combining the polarised light collected in the four sub-exposures, respectively. The Stokes $I$ spectra have a signal-to-noise (S/N) ratio ranging from 406 to 604 around 450 nm (see Table\,\ref{log}). 

\section{Magnetic field measurements}\label{magmeas}
\begin{figure*}
\begin{center}
\rotatebox{0}{\resizebox{\textwidth}{!}{\includegraphics{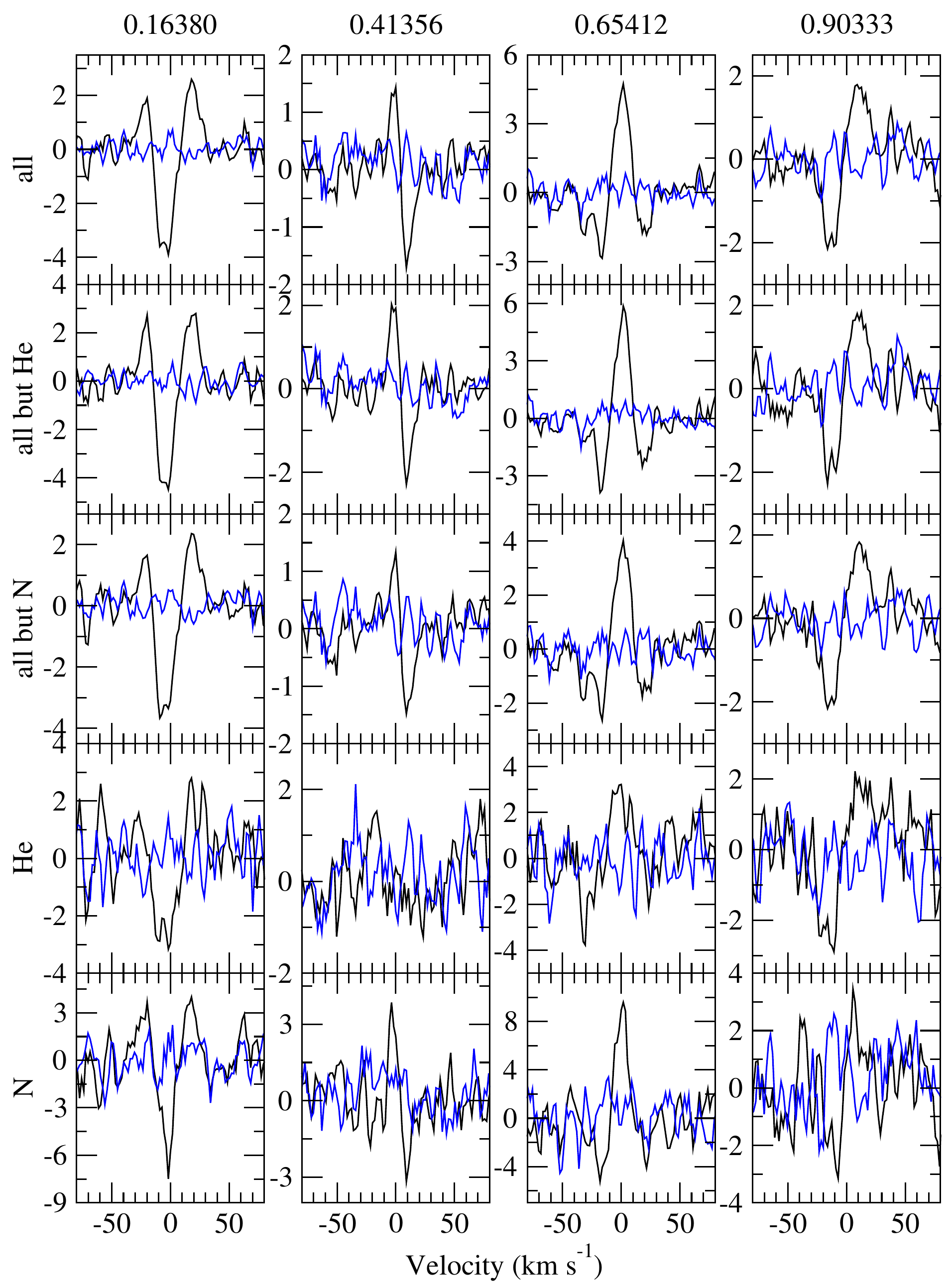}}}
\caption{Examples of LSD Stokes $V$ (black) and $N$ (blue) profiles taken at 4 typical rotational phases and for the 5 different line masks. The y-axis has been multiplied by 10$^4$.} 
\label{comp}
\end{center}
\end{figure*}

\subsection{Magnetic Zeeman signatures}

For each observation, we applied the Least-Squares Deconvolution (LSD) technique \citep{donati1997} to the photospheric spectral lines in each echelle spectrum (wavelength domain from 3750 to 10 500 \AA) in order to construct a single profile with an increased signal-to-noise. In that way, the Zeeman signature induced by the presence of a magnetic field is clearly visible in the high S/N LSD Stokes $V$ profiles, as seen in Figs.\,\ref{comp} and \ref{stokes}. To compute the LSD profiles, we made use of five different line masks created from the Kurucz atomic database and ATLAS~9 atmospheric models of solar abundance (Kurucz 1993). We used one mask containing 335 photospheric He and metal lines of various chemical elements (all lines hereafter), one mask with all 268 metal lines (all but He lines), one mask with all He and metal lines except the N lines (313 lines, all but N lines), one mask containing 67 He lines, and finally one mask containing 22 N lines. For each spectral line, the mask contains the wavelength, depth, and Land\'e factor, to be used by the LSD code. The depths were modified so that they correspond to the depth of the observed spectral lines. Lines whose Land\'e factor is unknown were excluded. The average S/N of the LSD Stokes $V$ profiles ranges from 5\,000 (mask with N lines only) to 14\,000 (mask with all lines) according to the number of lines used. 

In Fig.\,\ref{comp}, we compare LSD Stokes $V$ and $N$ profiles computed for different masks, for typical rotation phases. The absence of significant signatures in the null spectra of all our observations indicates that, thanks to our short exposures, none of our measurements has been polluted by stellar pulsations. All the LSD Stokes $V$ and $I$ profiles, computed for the mask with all but He lines, are displayed in Fig.\,\ref{stokes}. A magnetic Zeeman signature is present in almost all observations.

To statistically quantify the detection of a magnetic field in each of the Stokes $V$ profile, we used the False Alarm Probability (FAP) of a signature in the LSD Stokes $V$ profile inside the LSD line, compared to the mean noise level in the LSD Stokes $V$ measurement outside the line. We adopted the convention defined by \cite{donati1997}: if FAP $<$ 0.001\%, the magnetic detection is definite, if 0.001\% $<$ FAP $<$ 0.1\% the detection is marginal, otherwise there is no magnetic detection. The type of detection for each measurement is reported in Table~\ref{log_B}. We find that 74 of our 82 measurements exhibit a definite detection, while 3 measurements show a marginal detection, and there is no detection in 5 of them.

\subsection{Longitudinal magnetic field}\label{sect_Bl}

We used the center-of-gravity method on the LSD Stokes $V$ and $I$ profiles \citep{rees1979, wade2000} to compute the line-of-sight component of the magnetic field integrated over the visible stellar surface, which is called the longitudinal magnetic field and is given by 
\begin{equation}
B_l=(-2.14 \times 10^{11} G)\frac{\int v V(v) dv}{\lambda g c \int [1-I(v)]dv},
\end{equation}
where $\lambda$, in nm, is the mean, S/N-weighted wavelength, $c$ is the velocity of light in the same units as the velocity $v$, and $g$ is the mean, S/N-weighted value of the Land\'e factors of all lines used to construct the LSD profile. We used $\lambda$ = 482 nm and g = 1.22 for the mask including all lines, all but He lines, and all but N lines. We used $\lambda$ = 481 nm and g = 1.21 for the mask with He lines only, and $\lambda$ = 491 nm and g = 1.20 for the mask with N lines only. The integration limits were chosen large enough to cover the width of the line but also small enough to avoid artificially increased error bars due to noisy continuum. A range of 23 and 20 km~s$^{-1}$ around the line centre was adopted when including the He lines or not, respectively.

\begin{figure}
\begin{center}\
\rotatebox{0}{\resizebox{0.47\textwidth}{!}{\includegraphics{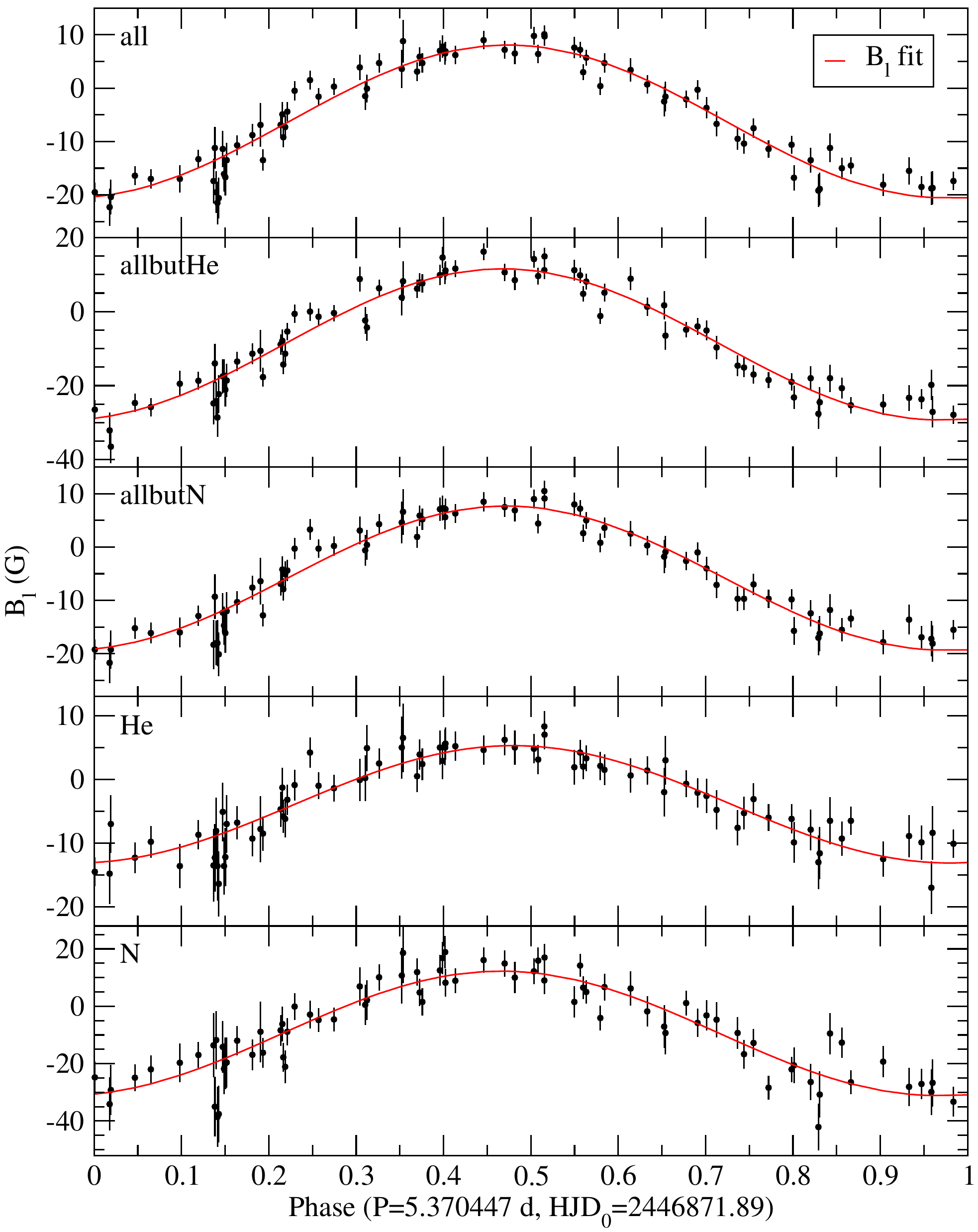}}}
\caption{Longitudinal magnetic field values plotted in phase with $P_{\rm rot}=5.370447$~d and HJD$_0 = 2446871.89$~d for the different groups of lines (from top to bottom: all photospheric He and metal lines, all photospheric lines except He lines, all photospheric lines except N lines, He lines only, and N lines only). A fit of the data for a dipole model (red line) is displayed in each panel. }  
\label{Bl}
\end{center}
\end{figure}

The values for the longitudinal magnetic field $B_l$ and null $N_l$ measurements are reported in Table\,\ref{log_B} for the different masks used. The $B_l$-values vary between about $-$25 and 10 G. The error bars are typically around 2.5, 3.5, 3.0, 3.1 and 6.5 G when using all, all but He, all but N, He, and N lines. It can be compared to the mean error bar of $\sim$29~G for the Musicos $B_l$-values from \cite{neiner2003b}. Phase diagrams of the longitudinal magnetic field for the rotation period are shown in Fig.\,\ref{Bl}. 

\section{Magnetic field modelling}\label{magmod}

\subsection{Longitudinal field modelling}\label{blmod}

Assuming a centered dipole field and following \cite{borra1980}, the magnetic obliquity angle $\beta$ is constrained by the observed ratio $B_{l\ \rm min}/B_{l\ \rm max} = \cos(i - \beta)/\cos(i + \beta)$, for a given stellar inclination angle $i$. The polar field $B_{\rm pol}$ is then deduced from $B_0 \pm B = 0.283 * B_{\rm pol} \cos(\beta \pm i)$, where $B_0$ and $B$ are the constant and the amplitude of a sine fit to the phase-folded $B_l$ values, respectively (red lines in Fig.~\ref{Bl}). 

We adopt an inclination angle $i$ of $30^\circ$ deduced from a modelling of the LSD Stokes profiles (see Sect.~\ref{stokesmod}), which is also obtained from our ZDI (Zeeman-Doppler Imaging) modelling presented in Sect.~\ref{ZDI}. The results of a dipole fit on the longitudinal magnetic field values are reported in Table\,\ref{Blfit}, considering the different groups of lines displayed in Fig.\,\ref{Bl}. For example, for the mask using all but He lines, we obtain $\beta = 104^\circ$ and $B_{\rm pol} = 149$~G, with an error of 1~deg and 5~G, respectively. We note that, even if the $B_0$ and $B$-values are slightly different when considering the different masks, we get almost the same ratio $B_{l\ \rm min}/B_{l\ \rm max}$ in all cases, so that the deduced $\beta$-value is the same for a given inclination angle $i$. What differs, depending on the adopted mask, is the $B_{\rm pol}$-value, which amounts to 158~G at maximum. Different chemical elements probe different layers of the star and have different distributions at the surfaces. Therefore, the measured longitudinal field value for a given element differs from another element, which translates into different $B_{\rm pol}$-values. 

In Table\,\ref{Blfit}, $\Delta\phi$ denotes the phase shift of the dipole fits compared to the ephemeris ($P_{\rm rot}=5.370447$~d and HJD$_0 = 2446871.89$~d) determined from the UV data by \cite{neiner2003b}. This ephemeris corresponds to the epoch of minimum in EW of UV C\,{\sc iv}, Si\,{\sc iv}, and N\,{\sc v} lines (see their Fig.\,2). Depending on the mask used, a phase shift between 0.020 and 0.033 is obtained. The errors of 0.054~d on HJD$_0$ and of 0.000078~d on the rotation period allow for a phase shift up to $\sim$0.07 between the old UV data and our recent Narval data, so that the dipole fits are compatible with the ephemeris derived from the UV data. 

\begin{table}
\centering
\caption{Results of a dipole fit on the longitudinal magnetic field values using the five groups of lines.  }
\begin{tabular}{ccccccc}
\hline\hline
	 & $B_0$ & $B$  & $\Delta\phi$  &$\beta$ & $B_{\rm pol}$\\
         & G     &  G  &    &  deg  & G   \\
\hline\\
all	 &   $-6.23$  &  14.32  & 0.027 & 104  & 104 \\
all but He & $-8.85$  &  20.42  & 0.033 & 104  & 149 \\ 
all but N  & $-5.80$  &  13.50  & 0.028 & 104  & 98  \\
He	 &   $-3.93$  &  9.20   & 0.020 & 104  & 67  \\
N	 &   $-9.44$  &  21.69  & 0.034 & 104  & 158 \\
[0.2cm] 
     
\hline
\end{tabular}
\tablefoot{The stellar inclination angle is fixed to 30 degrees, as deduced from our forward LSD Stokes modelling (Sect.~\ref{stokesmod}) and from our ZDI modelling (Sect.~\ref{ZDI}). The errors are of the order of 1~deg and 5~G, for  $\beta$ and $B_{\rm pol}$, respectively. }
\label{Blfit}
\end{table}

\begin{table*}
\centering
\caption{Magnetic field configuration obtained by modelling LSD Stokes profiles, for a mask including all but He lines, with a centred or off-centred oblique dipole magnetic field.}
\begin{tabular}{cccccc}
\hline\hline
	$i$ &	$\beta$	& $B_{\rm pol}$ &	$\Delta \Psi$ &	$d$ &	$\chi^2$\\
   deg &   deg     &  G          &               & R   &  \\
\hline
	30.71$^{0.75}_{0.76}$ &	107.09$^{0.51}_{0.52}$	&	98.19$^{0.75}_{0.75}$	&	0.0785$^{0.0012}_{0.0012}$	&	0		&	1.80649\\[0.2cm] 
	29.29$^{0.74}_{0.75}$ &	102.19$^{0.58}_{0.58}$	&	100.74$^{0.79}_{0.79}$	&	0.0751$^{0.0012}_{0.0012}$	&	0.0716$^{0.0035}_{0.0035}$ &	1.69504\\[0.2cm] 
\hline
\end{tabular}
\tablefoot{The columns give the parameter values of the best fitting models for the inclination angle $i$, the obliquity angle $\beta$, the polar field strength $B_{\rm pol}$, the phase shift between the model and the phases derived from the published UV ephemeris $\Delta \Psi$, the off-centring of the field along the magnetic axis $d$, and the reduced $\chi^2$ values.}
\label{param}
\end{table*}

\begin{figure*}
\begin{center}
\rotatebox{0}{\resizebox{\textwidth}{!}{\includegraphics{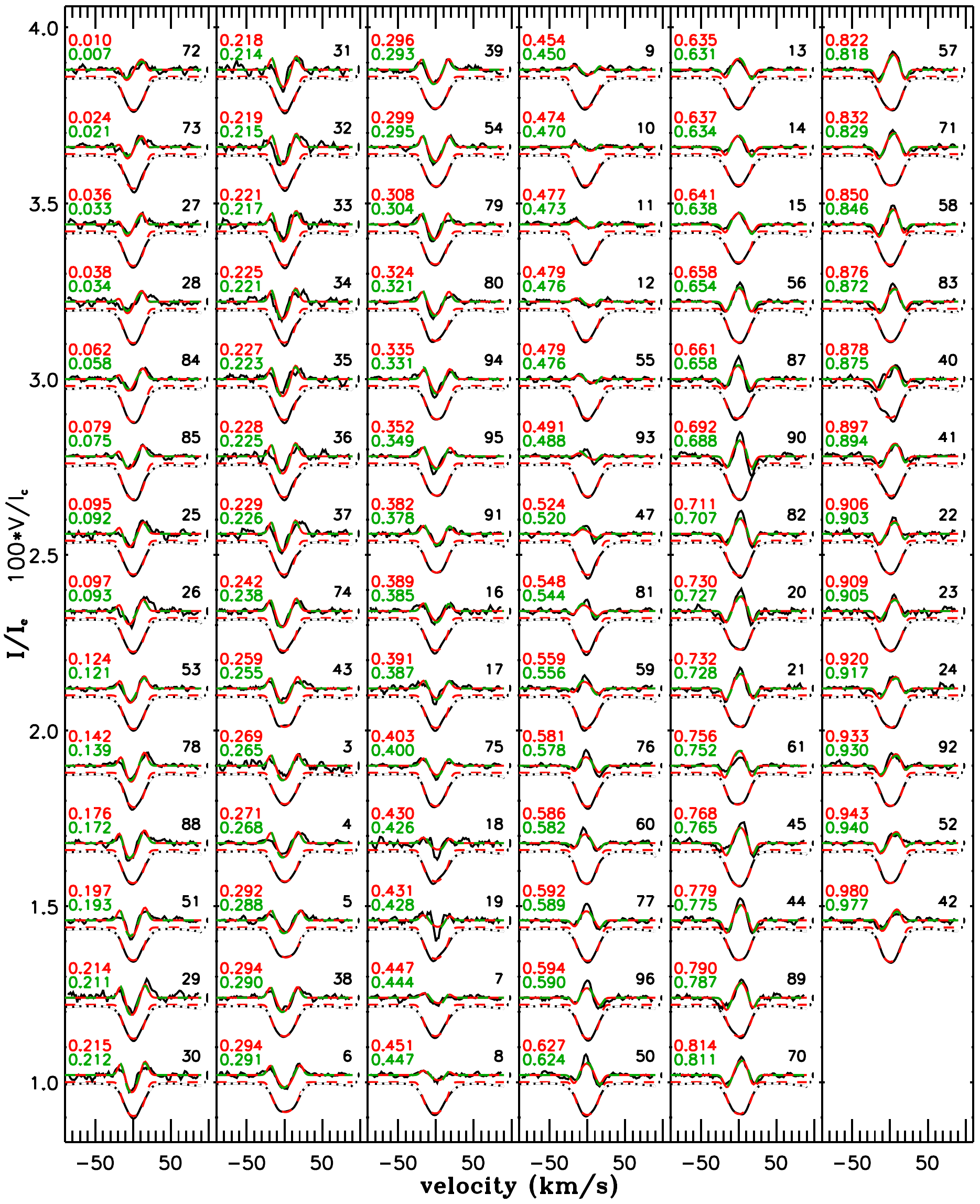}}}
\caption{Centred (solid red line) and off-centred (dashed green line) dipole field models superimposed on the LSD Stokes $V$ profiles (thin solid black lines) for all phases of observation and using all but He lines, and the corresponding modelled (dashed red line) and observed (dotted black line) LSD Stokes $I$ profiles. Above each profile, the rotation phase is indicated on the left and the number of the polarimetric sequence is indicated on the right. Typical error bars are shown next to each Stokes $V$ profile.} 
\label{stokes}
\end{center}
\end{figure*}

\subsection{Modelling the LSD Stokes profiles}\label{stokesmod}

We also determined the magnetic configuration of $\zeta$~Cas by fitting synthetic LSD Stokes $V$ and $I$ profiles to the observed ones for a centred dipole oblique rotator model. The free parameters of the model are the stellar inclination angle $i$ (in degrees), the magnetic obliquity angle $\beta$ (in degrees), and the dipole magnetic intensity $B_{\rm pol}$ (in Gauss). Afterwards, we checked whether a better fit is obtained with a dipole off-centred along the magnetic axis by a distance $d$ (in stellar radius). $d = 0$ corresponds to a centred dipole and $d = 1$ corresponds to a centre of the dipole at the surface of the star.

In our model, we use Gaussian local intensity profiles with a width calculated using the resolving power of Narval, a macroturbulence value of 2 km s$^{-1}$, and a $v\sin i$ of 20 km\,s$^{-1}$. The depth is determined by fitting the observed LSD Stokes $I$ profiles. The local Stokes $V$ profiles are calculated assuming the weak-field case and using the weighted mean Land\'e factor and wavelength of the lines derived from the LSD mask (see Sect.\,\ref{sect_Bl}). The synthetic Stokes $V$ profiles are obtained by integrating over the visible stellar surface by using a linear limb darkening law with a parameter equal to 0.3. The Stokes $V$ profiles are then normalised to the intensity continuum. 

 By varying the free parameters mentioned above, we calculated a grid of Stokes $V$ profiles for each phase of observation for the rotation period $P_{\rm rot}=5.370447$~d and the reference date HJD$_0 = 2446871.89$ corresponding to the time of minimum equivalent width of the wind-sensitive UV line, used in \cite{neiner2003b}. In our fitting procedure, a phase shift compared to the UV ephemeris $\Delta \Psi$ is allowed. We applied a $\chi^2$-minimization to determine the model that best matches our 82 observed profiles simultaneously. The fitting was applied to observed LSD profiles obtained with all but He lines. We did not model the LSD Stokes profiles for the masks including He lines because a Gaussian intrinsic profile is not suited in those cases. The fitting to LSD profiles computed with the nitrogen lines led to similar parameter values as when using all but He lines but with larger error bars, as expected from the lower S/N.

 The parameter values obtained for a centred and off-centred model are given in Table\,\ref{param}. The polar field strength obtained is of the order of 100~G, $i$ is about 30$^\circ$, and $\beta$ around 105$^\circ$. The values of the phase shift $\Delta\Psi$ of $<$0.08 remains compatible with the ephemeris deduced from the UV data. The errors on the model parameters were determined with the help of the MINUIT library, a physics analysis tool for function minimization developed at CERN\footnote{ http://seal.web.cern.ch/seal/snapshot/work-packages/mathlibs/minuit/}. The algorithm consists in varying a parameter at a time, the 3 or 4 other parameters being held fixed, to find the two values of the parameter for which the $\chi^2$ function is equal to its minimum (with respect to all parameters) plus a constant equal to 9 for 3$\sigma$ errors. This algorithm takes into consideration the non-linearities of the $\chi^2$ function with respect to its parameters, i.e. the deviation of the $\chi^2$ profile from a perfect parabola at its minimum. Therefore, it provides slightly asymmetrical errors (not visible within our precision however). We checked that determining the errors for each parameter independently is justified by computing the Hessian matrix. We indeed find that the correlation terms are small.

 The small value of the decentring parameter $d$ along with the only slightly better reduced $\chi^2$-values for the off-centred cases imply that our data are fully consistent with a dipolar magnetic field at the centre of the star. Figure\,\ref{stokes} shows the comparison between the observed and synthetic LSD V and I profiles for a centred and off-centred dipole for all phases of observation.

\subsection{Zeeman-Doppler Imaging }\label{ZDI}

\begin{figure}
\vspace{-0.8cm}
\rotatebox{0}{\resizebox{0.5\textwidth}{!}{\includegraphics{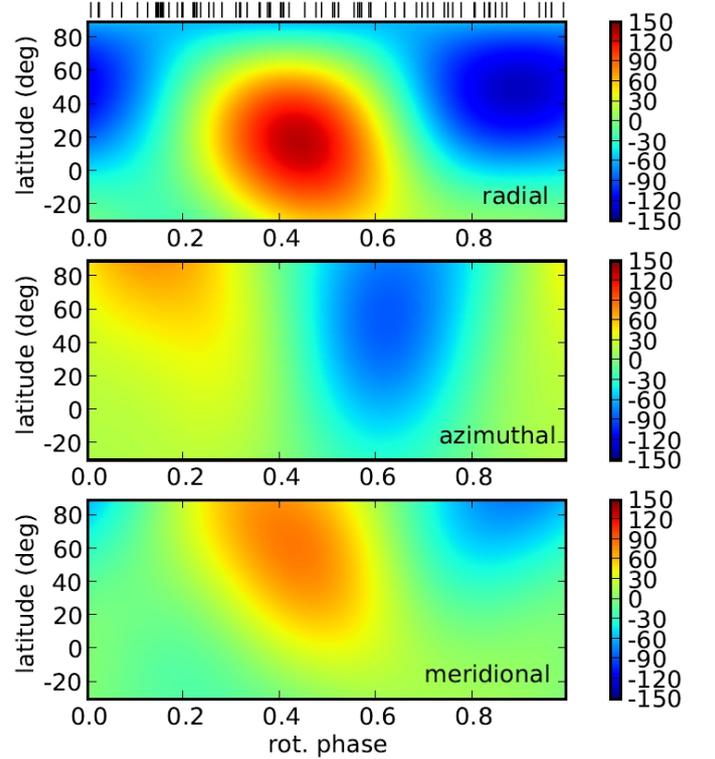}}}
\vspace{-1cm}
\caption{Magnetic map of $\zeta$~Cas. The three panels illustrate the field components in spherical coordinates (from top to bottom: radial, azimuthal, meridional). The magnetic field strength (colour scale) is expressed in Gauss. The vertical ticks on top of the radial map show the phases of observations. }  
\label{map}
\end{figure}

\begin{figure}
\vspace{-0.3cm}
\rotatebox{0}{\resizebox{0.55\textwidth}{!}{\includegraphics{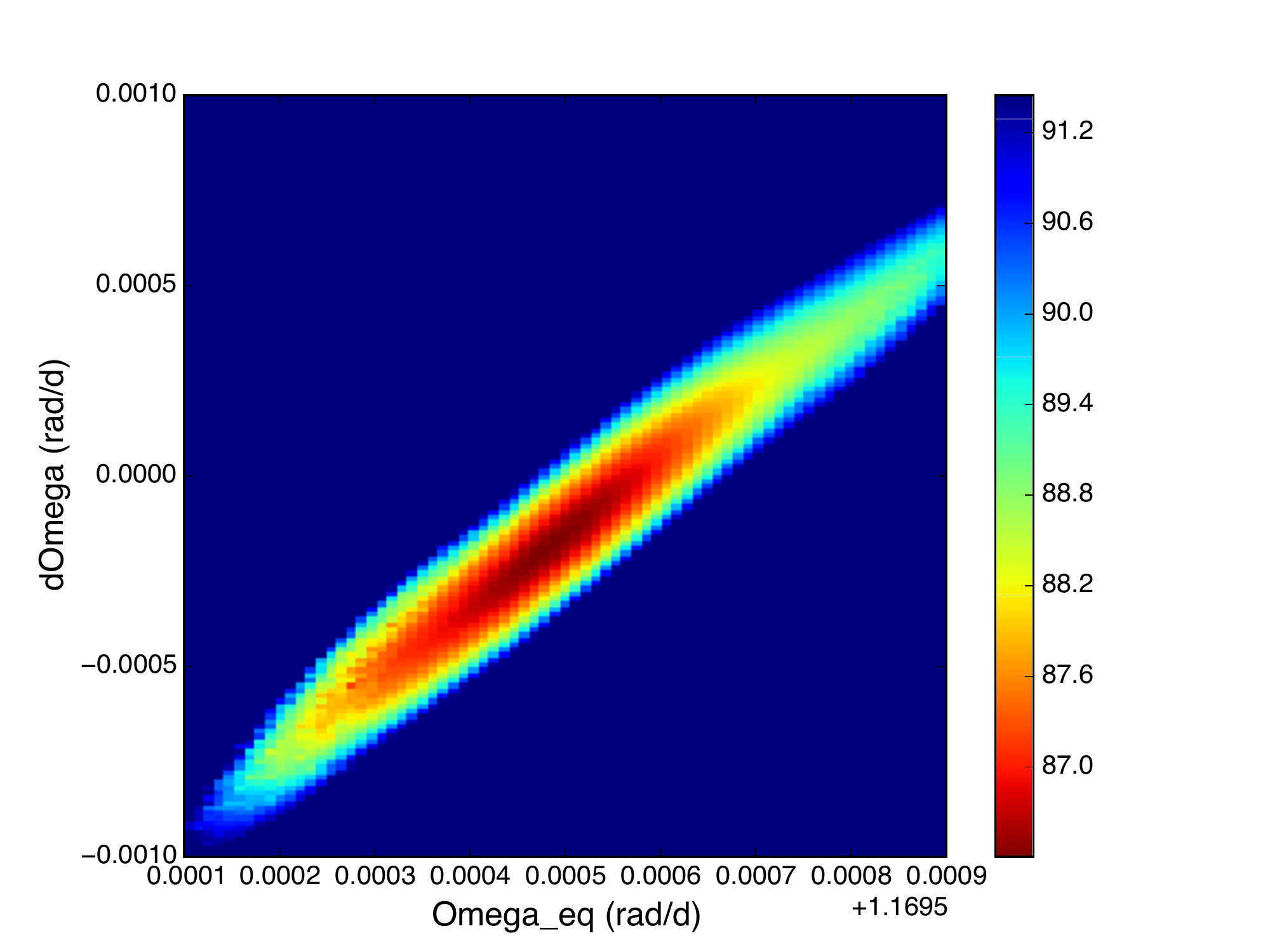}}}
\caption{Map of the mean surface magnetic field in Gauss in the differential rotation parameter plane.}  
\label{drot}
\end{figure}

To determine the surface map of the magnetic field we also used the Zeeman-Doppler Imaging technique (ZDI, \citealp{donati1997b}) following the procedure presented in \cite{petit2010} and using the LSD Stokes profiles for all but He lines. In this procedure, the model assumes a constant Stokes $I$ profile, which is not the case of $\zeta$~Cas. To avoid artefacts in the magnetic map due to the deformed intensity profiles, the spherical harmonic expansion was restricted to $\ell < 3$. The best model has a reduced $\chi^2$ of 1.83 and is obtained for a stellar inclination of 30$^{\circ}$. The corresponding ZDI map is shown in Fig.\,\ref{map}: the radial, azimuthal and meridional components are displayed. We find that the field has a simple configuration, compatible with an oblique dipole field, as expected for this kind of stars. The mean field strength at the surface amounts to 82~G and the maximum field strength is 148~G at a latitude of 23$^{\circ}$.

Thanks to our very good phase coverage and our long timebase of observations, we can measure the surface differential rotation of the star with high accuracy. To this end we used the method developed by \cite{petit2002}, assuming a simplified solar rotation law $\Omega(\ell) = \Omega_{\rm eq} - \rm{d}\Omega \sin^2 \ell$, where $\Omega(\ell)$ is the rotation rate at latitude $\ell$, $\Omega_{\rm eq}$ is the rotation rate of the equator and d$\Omega$ is the difference in rotation rate between the pole and equator. As shown in Fig.\,\ref{drot}, we obtain $\Omega_{\rm eq} = 1.16999$$\pm$$0.00002\ \rm{rad\ d}^{-1}$ and d$\Omega = -0.0002$$\pm$$0.0004\ \rm{rad\ d}^{-1}$. d$\Omega$ is found to be compatible with zero, which means that our Narval data are fully compatible with a solid body rotation at the surface, given the very low error bar. If differential rotation is present at a level below our detection threshold, our result is in favour of an anti-solar differential rotation, i.e. the pole would rotate faster than the equator.

\subsection{Comparison of the modelling results}

The best ZDI model is obtained for a stellar inclination of 30$^{\circ}$, in perfect agreement with the value derived from our forward modelling (Sect.\,\ref{stokesmod}). In addition, the $\beta$ angle (105$^{\circ}$) obtained from the $B_l$ values is similar to that deduced from the LSD Stokes modelling. However, the $B_{\rm pol}$ value found from fitting the LSD Stokes profiles (around 100~G, see Table\,\ref{param}) is lower than the value obtained from the $B_l$ values or from the ZDI models, which are both compatible (around 150~G), but one can consider an uncertainty of a factor of 2 on the magnetic strength reconstructed through ZDI \citep{vidotto2014}. 

For their magnetic field modelling, \cite{neiner2003b} used an inclination angle $i$ of $18^{\circ}$$\pm$$6^{\circ}$ while we obtain $i$$\sim$$30^{\circ}$. Their value was computed from $P_{\rm rot}$, $v\sin i=17$$\pm$$3$ km\,s$^{-1}$ and a stellar radius of $R=5.9$$\pm$$0.7R_\odot$. The inclination angle adopted in \cite{neiner2003b} is underestimated, because their $v\sin i$ value was underestimated. Our spectra of higher precision led to $v\sin i=20$$\pm$$3$ km\,s$^{-1}$, also obtained by \cite{nieva2012}, which is compatible with our value of $i$$\sim$$30^{\circ}$. 

We conclude that a dipole oblique rotator model with $i$$\sim$$30^{\circ}$, $\beta$$\sim$$105^{\circ}$, and $B_{\rm pol}$$\sim$100-150~G reproduces our spectropolarimetric measurements. 

\section{Discussion}\label{disc}

\subsection{Weak magnetic fields among B-type stars}

Current spectropolarimetric data give us new insights about main-sequence stars with radiative envelopes (see \citealp{briquet2015} for a review). Among intermediate-mass stars, there is evidence for two classes of magnetism: the Ap-like ``strong'' magnetism and the Vega-like ``ultra-weak'' magnetism. For the first group, the observed field is always higher than a threshold of 100~G for the longitudinal field, which corresponds to a polar field strength higher than  300~G \citep{auriere2010}. In the second group, sub-Gauss longitudinal fields have been convincingly detected (\citealp{lignieres2009}; \citealp{petit2010}; \citealp{blazere2016}). Between both groups, no star is observed and one speaks of a ``magnetic desert'' \citep{lignieres2014}.

This raises the question whether this dichotomy also holds for higher mass stars of spectral type O and B. For the latter, there are Bp stars, an extension of the Ap stars, i.e. stars with a typical polar strength of the order of kiloGauss, but no ``ultra-weak'' field has been observed so far. In fact, for massive stars, detecting fields with a longitudinal component below 1 Gauss requires the co-addition of even more high-resolution spectropolarimetric data than for A stars to reach the necessary signal-to-noise. This has only been attempted for two stars so far: $\gamma$\,Peg \citep{neiner2014b} and $\iota$\,Her \citep{wade2014}, so that further investigation to search for Vega-like fields in OB stars is being performed \citep{neiner2014c}. Besides that, evidence for fields with intermediate polar strength of the order of tens or a couple hundreds Gauss (``weak'' fields) is increasingly being observed in OB stars. So far, such weak fields have been observed in HD~37742 ($\zeta$ OriAa) (\citealp{blazere2015b}; \citealp{blazere2015a}), $\tau$~Sco \citep{donati2006}, $\epsilon$~CMa \citep{fossati2015}, and $\beta$~CMa \citep{fossati2015}. However, most of these targets are rather special: $\zeta$~OriAa is a young O supergiant and $\epsilon$~CMa is an evolved B star at the end of the main sequence. The increased radius of evolved stars explains their weak field compared to main-sequence stars, because of magnetic flux conservation. In addition, $\tau$~Sco has a complex field, therefore the disk-integrated field is weaker because of cancellation effects on the visible surface hemisphere. $\beta$~CMa seems to be a more normal B star, but it has not been studied in detail yet, so that its polar field strength is not clearly known. Our study of $\zeta$~Cas thus provides the first definite case of a weak polar field in a normal B star, based on a large amount of data allowing a modelling of its magnetic field configuration. As already pointed out in \cite{fossati2015}, our finding indicates a possible lack of a ``magnetic desert'' in massive stars, contrary to what is observed in main-sequence A-type stars, or a field strength threshold in massive stars lower than the one observed in intermediate-mass stars. In any case, $\zeta$~Cas exhibits the lowest polar field observed so far in a main-sequence B-type star.

\subsection{Evidence for solid-envelope rotation}
As discussed in \cite{mathis2015} and \cite{neiner2015}, there is increasing theoretical and observational evidence that the magnetic fields observed at the surface of intermediate and high-mass stars are remnants of an early phase of the life of the star. They are called fossil fields and reside inside the star without being continuously renewed. Furthermore, these magnetic fields deeply modify the transport of angular momentum and the mixing of chemicals. Above a critical field limit, the fields enforce a uniform rotation along field lines and the mixing is inhibited (\citealp{mathis2005}; \citealp{zahn2011}). The inhibition of mixing in the radiative zone by a magnetic field has been observed in the $\beta$~Cep star V2052~Oph using the combination of spectropolarimetry and asteroseismology (\citealp{neiner2012}; \citealp{briquet2012}). For the latter, the field strength observed ($B_{\rm pol}$$\sim$400~G) is 6 to 10 higher than the critical field limit needed to inhibit mixing as determined from theory. 

As in \cite{briquet2012}, we determined the critical field strength at the surface $B_{\rm crit,surf}$ for $\zeta$~Cas above which mixing is suppressed. In the criterion by \cite{zahn2011}, one makes use of the time already spent by $\zeta$~Cas on the main sequence $t$, its radius $R$ and mass $M$, as well as the surface rotation velocity. With $i=30^{\circ}$$\pm$$1^{\circ}$, $v \sin i = 20$$\pm$$3$ km\,s$^{-1}$ and $P_{\rm rot}=5.37045$~d, we have $v_{\rm eq} = 40$$\pm$$7.5$ km\,s$^{-1}$ and $R = 4.2$$\pm$$0.8 R_\odot$. To estimate $t$ and $M$, we computed main-sequence stellar models accounting for $T_{\rm eff} = 20750$$\pm$$200$ K and $\log g = 3.80$$\pm$$0.05$ dex \citep{nieva2012}, within 3$\sigma$. To this end, we used the evolutionary code CL\'ES (Code Li\'egeois d'\'Evolution Stellaire; \citealp{scuflaire2008}) with the input physics described in \cite{briquet2011}. The hydrogen mass fraction $X$ is set to 0.7. Allowing a core overshooting parameter value $\alpha_{\rm ov}$ between 0 and 0.4 (expressed in local pressure scale heights) and a metallicity between 0.01 and 0.02, we obtain $M \in [6.7,9.5] M_\odot$, $R \in [4.5,7.6] R_\odot$, and $t \in [19,40]$ Myr. If we limit to $R \in [4.5,5.0] R_\odot$ to be compatible with the observational constraints on the stellar radius, we have $M \in [6.7,8.0] M_\odot$, $t \in [22,40]$ Myr, and $\log g \in [3.88,3.95]$. Furthermore, if we set $\alpha_{\rm ov} = 0$, we get $M \in [7.1,8.0] M_\odot$, $t \in [22,29]$ Myr, and $\log g \in [3.89,3.95]$.

We found that the mean critical field strength in the radiative zone for suppressing differential rotation in this star is $B_{\rm crit} = 45$~G according to the criterion from Eq.~(3.6) in \cite{zahn2011}. According to the ratio (30) between internal and surface fields derived by \cite{braithwaite2008} (see Fig.\,8 therein), this corresponds to a critical field strength at the surface of $B_{\rm crit,surf} = 1.5$~G. 

Considering that the observed field strength is 100-150 G, we expect rigid envelope rotation and no core overshooting for $\zeta$~Cas. We point out that the adopted criterion by \cite{zahn2011} is applicable only in the case of a magnetic obliquity angle $\beta = 0$, while $\zeta$~Cas has a large obliquity. Therefore, our result needs to be confirmed whenever a similar critical field criterion taking the obliquity angle into account will be available (\citealp{neiner2015b}; Mathis et al., in prep.). Nevertheless, the conclusion still holds when considering another criterion from Eq.~(22) defined by \cite{spruit1999}, which does take obliquity into account and gives $B_{\rm crit,surf} < 35$~G. Therefore, it is likely that the envelope of the star rotates uniformly. Note that the above two criteria could also be influenced by the effect of waves, but this has never been studied. Therefore, we conclude that our study provides the first evidence for a solid body rotation in the envelope of a weak-field B-type star, but that this conclusion must be confirmed with more elaborated critical field criteria in the future.

\subsection{Surface N-enrichment in early B-type stars}

Early B-type stars are ideal indicators for present-day cosmic abundances \citep{nieva2012} but, in a fraction of them, the stellar surface is found to be enriched with the products from hydrogen burning, such as nitrogen (e.g., \citealp{morel2006}). This is the case of $\zeta$~Cas. An explanation to the surface enrichment of certain chemical species is that internal mixing can lead to the dredge up of some core-processed CNO material. The usual interpretation is that rotationally-induced mixing is responsible for the N-enrichment observed in massive stars \citep{meynet2000}. Such a mixing is enhanced when magnetic braking occurs in models with interior differential rotation \citep{meynet2011} and the predicted N-enrichments depend on the rotation velocity, but also on the mass, age, binarity, metallicity and magnetic field \citep{maeder2015}. As discussed in the previous section, we proved that the envelope of $\zeta$~Cas rotates as a solid body. Therefore, the nitrogen overabundance in its stellar surface cannot be explained by rotationally-induced mixing. This conclusion was also reached for the other magnetic N-rich $\beta$~Cep pulsator V2052~Oph. For the latter, an asteroseismic study showed that the stellar models reproducing the observed pulsational behaviour do not have any convective core overshooting and the observed fossil magnetic field is strong enough to inhibit the interior mixing and enforce rigid rotation in the envelope (\citealp{briquet2012}; \citealp{neiner2012}). 

A surface N-enhancement linked to binarity \citep{langer2008} is also ruled out by our observations for both $\zeta$~Cas and V2052~Oph. Therefore, yet another mechanism must be invoked to explain the abundance peculiarity. As discussed in \cite{aerts2014}, recent studies suggest that internal gravity waves might also play a role to account for the surface abundances. Indeed, theoretical and numerical simulations show that large-scale low-frequency internal gravity waves are excited by the convective core and can transport angular momentum (\citealp{mathis2013}; \citealp{rogers2013}). Such waves were revealed by high-precision space-based CoRoT photometric data in several Be \citep{neiner2012b} and O-type stars. For the latter, a red noise power excess of physical origin was detected by \cite{blomme2011} and then interpreted as internal gravity waves by \cite{aerts2015}. Clearly, such waves are present in massive stars and their transport modify the mixing in their radiative zone. Contrary to the case of solar-like stars (\citealp{press1981}; \citealp{schatzman1996}), the chemical transport that such waves induce in massive stars remains to be computed, so that a comparison between theoretical predictions and observations is currently not possible. 
However, our observational study of $\zeta$~Cas leads us to conclude that the most likely explanation for the nitrogen enrichment at its surface is the transport of chemicals from the core to the surface by internal gravity waves. 

Several authors (\citealp{shiode2013}; \citealp{rogers2013}; \citealp{lee2014}) have already demonstrated, thanks to analytical work or simulations, that the amplitude of these waves is sufficient to propagate to the surface and thus transport angular momentum and chemicals. Internal gravity waves propagate in the radiation zone and deposit or extract (depending on whether they are prograde or retrograde) angular momentum at the location where they are dissipated (\citealp{goldreich1989}; \citealp{zahn1997}). Assuming the convective core rotates faster than the radiative envelope, the net angular momentum transport leads to an acceleration of the surface layers \citep{lee2014}. In the case of $\zeta$~Cas, we showed that rotation is likely uniform in the envelope due to the presence of the magnetic field. However, the rotation of the core could be higher or lower than the rotation of the envelope, and waves could locally increase differential rotation just below the surface. 

Moreover, the magnetic field transforms the gravity waves into magneto-gravity waves. However, if the frequency of the waves are higher than the Alfven frequency, the properties of the waves are only slightly modified by a weak magnetic field, such as the one observed in $\zeta$~Cas (\citealp{mathis2012}; \citealp{mathisalvan2013}). In particular, radiative damping increases with the magnetic field strength: the angular momentum transport by waves thus becomes an inverse function of the field strength \citep{mathis2012}. However, the weak magnetic field of $\zeta$~Cas will not hamper much the transport by waves.

\section{Summary}\label{sum}

With the aim to better constrain the magnetic field configuration of $\zeta$~Cas, we made use of 82 Narval spectropolarimetric data well distributed over the accurately known rotation cycle of this N-rich magnetic B-type pulsator. Zeeman signatures typical of the presence of a magnetic field are seen in almost all LSD Stokes $V$ profiles, while the diagnosis null spectra do not show any signature. This confirms that the signatures are of magnetic origin.

To model the magnetic field, we considered a dipole oblique rotator model and we fitted both the longitudinal measurements and the LSD Stokes $V$ and $I$ profiles to derive the parameter values of the model. Besides this forward approach, we also used the Zeeman Doppler Imaging method to derive the surface map of the magnetic field of the star. Our magnetic modelling leads to a stellar inclination angle $i$ of  30$^\circ$, a magnetic obliquity angle $\beta$ of 105$^\circ$, and a polar field strength $B_{\rm pol}$ of 100-150~G. 

Although weak magnetic fields have already been reported for several massive stars, $\zeta$~Cas is the only firmly established case of a weak polar field in a normal B star and exhibits the lowest field observed so far in a massive main-sequence star. We also point out that $\zeta$~Cas has a polar field strength in the range where a so-called ``magnetic desert'' between about 1~G and 300~G is observed for intermediate-mass stars. This indicates that either there is no magnetic desert for massive stars or such a desert is less extended for massive stars than for intermediate-mass stars.

In addition, thanks to our intensive dataset, our ZDI modelling allowed us to test whether the star shows surface differential rotation. Besides that, a theoretical criterion was used to determine the critical field strength above which mixing is suppressed. Both approaches led to the conclusion that $\zeta$~Cas rotates as a solid body in the envelope. Indeed, our Narval data prove that there is no surface differential rotation, while theory predicts that the magnetic field enforced an interior rigid rotation in the radiative zone. 

To explain the nitrogen enrichment that is observed at the surface of a fraction of massive stars, and in $\zeta$~Cas, the proposed explanations are rotationally-induced mixing, a binarity origin, or a pulsational cause. For $\zeta$~Cas, the first two hypothesis are ruled out thanks to our study since interior differential rotation is excluded and there is no evidence of binarity in the high-resolution high S/N spectropolarimetry taken on a long time base. Therefore, the pulsational origin remains the most convincing explanation for the observed peculiarity. $\zeta$~Cas is likely N-rich due to the transport of core-processed CNO material from the core to the surface by internal gravity waves.

\begin{acknowledgements} 
The authors would like to thank St\'ephane Mathis for useful discussions. CN acknowledges support from the ANR (Agence Nationale de la Recherche) project
Imagine. This research has made use of the SIMBAD database operated at CDS,
Strasbourg (France), and of NASA's Astrophysics Data System (ADS).
\end{acknowledgements}

\bibliographystyle{aa}
\bibliography{zetacas_paper}

\appendix 
\section{Journal of Narval/TBL observations and magnetic field measurements}
\begin{table*}
  \caption{Journal of Narval/TBL observations of $\zeta$~Cas. Column~1/8 indicates the number of the polarimetric sequence. Columns~2/9 and 3/10 show the date and time of the middle of observations, while Col.~4/11 gives the Heliocentric Julian Date (HJD) at the middle of observations. Column~5/12 gives the exposure time in seconds for each sequence, and Col.~6/13 the rotational phase using $P_{\rm rot}=5.370447$~d and the reference date HJD$_0 = 2446871.89$. Column~7/14 give the signal-to-noise ratio of the Stokes $I$ spectra around 450 nm. A missing phase and S/N indicates that the corresponding observation was not used in the analysis (see Sect.\,\ref{obs} for explanations). }
\centering
\tiny
 \begin{tabular}{ccccccc|ccccccc}
\hline\hline
Nr. & Date  & Mid-UT &  Mid-HJD & $T_{\rm exp}$ & Phase & S/N &  Nr. & Date  & Mid-UT &  Mid-HJD & $T_{\rm exp}$ & Phase & S/N\\
 &   & h:min &  $-$2450000 & s & & & &  & h:min &  $-$2450000 & s & &\\
\hline
1 & 2007-01-11 & 17:56 & 4112.25359 & 4x180 & -        & - & 49 & 2011-07-12 & 02:22 & 5754.60870 & 4x420 & - & -\\
2 & 2007-01-11 & 18:12 & 4112.26442 & 4x180 & -        & - & 50 & 2011-07-15 & 01:15 & 5757.56212 & 4x420 & 0.54982 & 530\\
3 & 2007-01-11 & 18:27 & 4112.27525 & 4x180 & 0.19043  & 452 & 51 & 2011-07-18 & 02:39 & 5760.62105 & 4x420 & 0.11941 & 568\\
4 & 2007-01-11 & 18:43 & 4112.29052 & 4x360 & 0.19327  & 557 & 52 & 2011-07-22 & 02:56 & 5764.63256 & 4x420 & 0.86637 & 560\\
5 & 2007-01-11 & 21:22 & 4112.39931 & 4x300 & 0.21353  & 556 & 53 & 2011-07-23 & 02:11 & 5765.60078 & 4x380 & 0.04665 & 571\\
6 & 2007-01-11 & 21:45 & 4112.41551 & 4x300 & 0.21655  & 557 & 54 & 2011-08-09 & 03:18 & 5782.64961 & 4x420 & 0.22122 & 549\\
7 & 2007-01-12 & 17:30 & 4113.23806 & 4x300 & 0.36971  & 533 & 55 & 2011-08-10 & 02:38 & 5783.62187 & 4x420 & 0.40226 & 529\\
8 & 2007-01-12 & 17:53 & 4113.25427 & 4x300 & 0.37273  & 548 & 56 & 2011-08-11 & 01:29 & 5784.57366 & 4x420 & 0.57948 & 544\\
9 & 2007-01-12 & 18:16 & 4113.27047 & 4x300 & 0.37574  & 540 & 57 & 2011-08-11 & 22:41 & 5785.45747 & 4x420 & 0.74405 & 513\\
10 & 2007-01-12 & 20:52 & 4113.37863 & 4x300 & 0.39588 & 539 & 58 & 2011-08-12 & 02:19 & 5785.60837 & 4x420 & 0.77215 & 542\\
11 & 2007-01-12 & 21:15 & 4113.39484 & 4x300 & 0.39890 & 527 & 59 & 2011-08-15 & 21:45 & 5789.41837 & 4x420 & 0.48159 & 498\\
12 & 2007-01-12 & 21:39 & 4113.41103 & 4x300 & 0.40192 & 523 & 60 & 2011-08-16 & 01:11 & 5789.56182 & 4x420 & 0.50830 & 533\\
13 & 2007-01-13 & 17:32 & 4114.24066 & 4x360 & 0.55640 & 557 & 61 & 2011-08-16 & 23:02 & 5790.47176 & 4x420 & 0.67774 & 539\\
14 & 2007-01-13 & 17:59 & 4114.25966 & 4x360 & 0.55994 & 567 & 62 & 2011-08-18 & 01:42 & 5791.58315 & 4x420 & - & -\\
15 & 2007-01-13 & 18:26 & 4114.27865 & 4x360 & 0.56347 & 590 & 63 & 2011-08-18 & 23:10 & 5792.47749 & 4x420 & - & -\\
16 & 2007-01-17 & 18:48 & 4118.28952 & 4x180 & 0.31031 & 536 & 64 & 2011-08-19 & 03:11 & 5792.64514 & 4x420 & - & -\\
17 & 2007-01-17 & 19:04 & 4118.30017 & 4x180 & 0.31230 & 516 & 65 & 2011-08-20 & 00:22 & 5793.52760 & 4x420 & - & -\\
18 & 2007-06-28 & 03:01 & 4279.62780 & 4x120 & 0.35218 & 440 & 66 & 2011-08-21 & 00:21 & 5794.52718 & 4x420 & -   & -\\
19 & 2007-06-28 & 03:13 & 4279.63574 & 4x120 & 0.35366 & 423 & 67 & 2011-08-21 & 03:21 & 5794.65207 & 4x420 & -  & -\\ 
20 & 2007-07-05 & 02:37 & 4286.61181 & 4x120 & 0.65264 & 504 & 68 & 2011-08-22 & 01:09 & 5795.56073 & 4x420 & - & -\\
21 & 2007-07-05 & 02:49 & 4286.61975 & 4x120 & 0.65412 & 516 & 69 & 2011-08-22 & 23:08 & 5796.47638 & 4x420 & - & -\\
22 & 2007-07-06 & 01:23 & 4287.55989 & 4x120 & 0.82917 & 477 & 70 & 2011-08-28 & 00:23 & 5801.52906 & 4x420 & 0.73665 & 530\\
23 & 2007-07-06 & 01:34 & 4287.56783 & 4x120 & 0.83065 & 480 & 71 & 2011-08-28 & 02:44 & 5801.62710 & 4x420 & 0.75491 & 554\\
24 & 2007-07-06 & 03:06 & 4287.63152 & 4x120 & 0.84251 & 507 & 72 & 2011-08-29 & 01:38 & 5802.58321 & 4x420 & 0.93294 & 479\\
25 & 2007-07-07 & 01:41 & 4288.57244 & 4x120 & 0.01771 & 456 & 73 & 2011-09-24 & 23:56 & 5829.51186 & 4x420 & 0.94717 & 555\\
26 & 2007-07-07 & 01:52 & 4288.58039 & 4x120 & 0.01919 & 454 & 74 & 2011-09-26 & 03:52 & 5830.67530 & 4x420 & 0.16380  & 549\\
27 & 2007-07-12 & 02:55 & 4293.62469 & 4x120 & 0.95846 & 514 & 75 & 2011-09-27 & 00:47 & 5831.54736 & 4x420 & 0.32619 & 569\\
28 & 2007-07-12 & 03:07 & 4293.63261 & 4x120 & 0.95994 & 503 & 76 & 2011-09-27 & 23:39 & 5832.49977 & 4x420 & 0.50353 & 601\\
29 & 2007-07-13 & 01:54 & 4294.58223 & 4x120 & 0.13676 & 406 & 77 & 2011-09-28 & 01:10 & 5832.56337 & 4x420 & 0.51537 & 587\\
30 & 2007-07-13 & 02:06 & 4294.59018 & 4x120 & 0.13824 & 427 & 78 & 2011-10-01 & 00:00 & 5835.51466 & 4x420 & 0.06491 & 567\\
31 & 2007-07-13 & 02:17 & 4294.59811 & 4x120 & 0.13972 & 433 & 79 & 2011-10-01 & 21:13 & 5836.39858 & 4x420 & 0.22950 & 578\\
32 & 2007-07-13 & 02:28 & 4294.60603 & 4x120 & 0.14119 & 422 & 80 & 2011-10-01 & 23:30 & 5836.49369 & 4x420 & 0.24721 & 582\\
33 & 2007-07-13 & 02:40 & 4294.61396 & 4x120 & 0.14267 & 441 & 81 & 2011-10-03 & 04:13 & 5837.69019 & 4x420 & 0.47001 & 577\\
34 & 2007-07-13 & 03:15 & 4294.63865 & 4x120 & 0.14727 & 488 & 82 & 2011-10-04 & 01:14 & 5838.56638 & 4x420 & 0.63316 & 604\\
35 & 2007-07-13 & 03:27 & 4294.64658 & 4x120 & 0.14874 & 480 & 83 & 2011-10-04 & 22:32 & 5839.45397 & 4x420 & 0.79843 & 504\\
36 & 2007-07-13 & 03:39 & 4294.65479 & 4x120 & 0.15027 & 470 & 84 & 2011-10-05 & 22:26 & 5840.44932 & 4x420 & 0.98377 & 574\\
37 & 2007-07-13 & 03:50 & 4294.66271 & 4x120 & 0.15175 & 469 & 85 & 2011-10-06 & 00:38 & 5840.54124 & 4x420 & 0.00088 & 588\\
38 & 2007-11-24 & 18:07 & 4429.26647 & 4x300 & 0.21554 & 566 & 86 & 2011-10-06 & 04:27 & 5840.70051 & 4x420 & - & - \\
39 & 2007-11-24 & 18:31 & 4429.28320 & 4x300 & 0.21865 & 578 & 87 & 2011-10-09 & 03:49 & 5843.67386 & 4x420 & 0.58419 & 546\\
40 & 2007-11-27 & 21:36 & 4432.41177 & 4x300 & 0.80121 & 532 & 88 & 2011-10-11 & 22:03 & 5846.43368 & 4x420 & 0.09808 & 494\\
41 & 2009-09-23 & 22:27 & 5098.44989 & 4x420 & 0.82033 & 507 & 89 & 2011-10-25 & 23:02 & 5860.47453 & 4x420 & 0.71255 & 547\\
42 & 2009-10-15 & 20:42 & 5120.37741 & 4x420 & 0.90333 & 523 & 90 & 2011-10-30 & 19:14 & 5865.31675 & 4x420 & 0.61419 & 558\\
43 & 2009-10-28 & 02:24 & 5132.61064 & 4x420 & 0.18121 & 485 & 91 & 2011-11-08 & 21:04 & 5874.39264 & 4x420 & 0.30416 & 502\\
44 & 2009-11-26 & 17:46 & 5162.25461 & 4x420 & 0.70104 & 529 & 92 & 2011-11-11 & 20:11 & 5877.35603 & 4x420 & 0.85595 & 532\\
45 & 2011-07-05 & 01:39 & 5747.57841 & 4x420 & 0.69081 & 583 & 93 & 2011-11-30 & 22:44 & 5896.46199 & 4x420 & 0.41356 & 589\\
46 & 2011-07-06 & 02:01 & 5748.59145 & 4x300 & -       & - & 94 & 2011-12-10 & 20:20 & 5906.36150 & 4x420 & 0.25690 & 555\\ 
47 & 2011-07-09 & 02:58 & 5751.63355 & 4x420 & 0.44590 & 538 & 95 & 2011-12-10 & 22:37 & 5906.45664 & 4x420 & 0.27461 & 559\\
48 & 2011-07-11 & 03:02 & 5753.63654 & 4x420 & -       & - & 96 & 2012-01-18 & 19:59 & 5945.34444 & 4x420 & 0.51568 & 461\\
\hline
\end{tabular}
\label{log}
\end{table*}

\newpage

{\tiny
\begin{longtable}{cccccccccccc}
  \caption{Magnetic field measurements of $\zeta$~Cas. Column~1 indicates the number of the polarimetric sequence. The following 5 columns provide the longitudinal magnetic field value $B_l$ in Gauss, for LSD using all photospheric lines (Col.~2), using all photospheric lines but He lines (Col.~3), using all photospheric lines but N lines (Col.~4), using He lines only (Col.~5), and using N lines only (Col.~6). Columns~7 to 11 give the null measurements in Gauss for the five same masks. The last column gives the detection status of a magnetic field in Stokes $V$, where ND, MD, and DD, mean ``no detection'', ``marginal detection'', and ``definite detection'', respectively. }\\
\hline\hline
Nr. & $B_l$ & $B_l$ & $B_l$  &$B_l$  &  $B_l$  & $N_l$  & $N_l$   &$N_l$   &$N_l$  & $N_l$   & Detection\\
& (all) & (all but He) &  (all but N) & (He) & (N)& (all) & (all but He) &  (all but N) & (He) & (N)  & \\
   & G & G & G & G & G & G & G & G & G & G  & \\
\hline
\endfirsthead
\caption{continued.}\\  
\hline\hline
 Nr. & $B_l$ & $B_l$ & $B_l$  &$B_l$  &  $B_l$  & $N_l$  & $N_l$   &$N_l$   &$N_l$  & $N_l$  & Detection\\
& (all) & (all but He) &  (all but N) & (He) & (N)& (all) & (all but He) &  (all but N) & (He) & (N)  & \\
   & G & G & G & G & G & G & G & G & G & G  & \\
\hline
\endhead
\hline
3 & -6.9$\pm$4.1 & -10.6$\pm$5.6 & -6.4$\pm$4.4 & -7.8$\pm$5.2 & -8.9$\pm$10.6 & -1.8$\pm$4.1 & -4.6$\pm$5.6  & -2.8$\pm$4.4 & -0.7$\pm$5.3 & -17.8$\pm$10.9 & MD\\
4 & -13.5$\pm$2.0 & -17.7$\pm$2.6 & -12.8$\pm$2.1 & -8.5$\pm$2.7 & -16.2$\pm$5.2 & -0.2$\pm$2.0  &  -1.8$\pm$2.6 &  1.1$\pm$2.1 & 2.8$\pm$2.7 & -6.7$\pm$5.2 & DD\\
5 & -6.9$\pm$2.1 & -8.9$\pm$2.8 & -6.9$\pm$2.2 & -4.7$\pm$2.8 & -8.4$\pm$5.4 & -0.3$\pm$2.1 &  -0.8$\pm$2.8 & -0.8$\pm$2.2 & 1.4$\pm$2.8  & 2.3$\pm$5.4 & DD\\
6 & -9.2$\pm$1.9 & -14.3$\pm$2.6 & -7.9$\pm$2.1 & -5.8$\pm$2.6 & -17.8$\pm$5.0 & 0.5$\pm$1.9 & -2.3$\pm$2.6 & 1.5$\pm$2.1 & 2.7$\pm$2.6 & -2.5$\pm$5.0 & DD\\
7 & 3.1$\pm$1.8 & 6.2$\pm$2.5 & 1.9$\pm$2.0 & 0.5$\pm$2.5 & 11.9$\pm$4.8 & -0.8$\pm$1.8 &  1.2$\pm$2.5  & 0.5$\pm$2.0 & 1.4$\pm$2.5 & -9.5$\pm$4.8  & DD\\
8 & 5.9$\pm$1.8 & 7.9$\pm$2.5 & 5.9$\pm$1.9 & 3.9$\pm$2.4 & 4.8$\pm$4.8 & -0.7$\pm$1.8 & -3.4$\pm$2.5 & -0.4$\pm$1.9  & 0.0$\pm$2.4 & -2.9$\pm$4.8 & DD \\
9 & 4.7$\pm$1.8 & 7.6$\pm$2.5 & 5.2$\pm$2.0 & 2.4$\pm$2.5 & 1.5$\pm$4.8 & -1.1$\pm$1.8  & -0.7$\pm$2.5 & -1.3$\pm$2.0 & 1.2$\pm$2.5 & 4.8$\pm$4.8  & DD\\
10 & 7.0$\pm$2.0 & 9.9$\pm$2.8 & 7.1$\pm$2.2 & 5.0$\pm$2.7 & 12.5$\pm$5.4 & 0.3$\pm$2.0 &  2.0$\pm$2.8 & -0.1$\pm$2.2 & -1.0$\pm$2.7  & -1.5$\pm$5.4 & ND\\
11 & 7.8$\pm$2.1 & 14.6$\pm$2.9 & 7.3$\pm$2.3 & 2.9$\pm$2.9 & 16.8$\pm$5.6 & -1.4$\pm$2.1 & -0.8$\pm$2.9  & -0.4$\pm$2.3 & -0.7$\pm$2.9 & -2.2$\pm$5.6 & ND\\
12 & 6.5$\pm$2.1 & 10.3$\pm$2.9 & 5.6$\pm$2.3 & 5.0$\pm$2.8 & 18.9$\pm$5.6 & 0.8$\pm$2.1 & 3.4$\pm$2.9 &  0.7$\pm$2.3 & -0.9$\pm$2.8 & 4.8$\pm$5.6 & ND \\
13 & 7.2$\pm$1.5 & 9.8$\pm$2.1 & 7.2$\pm$1.6 & 4.2$\pm$2.0 & 14.2$\pm$4.0 & 0.8$\pm$1.5 & 2.2$\pm$2.1 & 0.4$\pm$1.6  & -0.2$\pm$2.0 & 3.5$\pm$4.0 & DD\\
14 & 3.0$\pm$1.5 & 4.8$\pm$2.1 & 2.6$\pm$1.6 & 2.0$\pm$2.0 & 6.5$\pm$4.0 & 2.7$\pm$1.5 & 4.3$\pm$2.1 &  2.4$\pm$1.6 & 1.3$\pm$2.0 & 9.1$\pm$4.0 &  DD\\
15 & 5.7$\pm$1.5 & 8.1$\pm$2.1 & 5.0$\pm$1.6 & 3.3$\pm$2.0 & 5.0$\pm$4.0 & -1.8$\pm$1.5 & -4.6$\pm$2.1 &  -1.7$\pm$1.6 & -0.7$\pm$2.0 & -0.2$\pm$4.0 & DD \\
16 & -1.5$\pm$2.6 & -2.4$\pm$3.6 & -0.6$\pm$2.9 & 0.2$\pm$3.6 & 0.5$\pm$7.0 & -0.7$\pm$2.6 & 1.1$\pm$3.6 &  -1.3$\pm$2.9 & 0.7$\pm$3.6 & -0.7$\pm$7.0 & DD\\
17 & -0.1$\pm$2.6 & -4.3$\pm$3.6 & 0.4$\pm$2.8 & 4.9$\pm$3.6 & 2.1$\pm$6.9 & 1.0$\pm$2.6 & -1.9$\pm$3.6 &  2.2$\pm$2.8 & 3.0$\pm$3.6 & -8.9$\pm$6.9 & DD\\
18 & 3.6$\pm$3.6 & 3.8$\pm$4.9 & 4.6$\pm$3.9 & 5.0$\pm$4.9 & 10.7$\pm$9.7 & -3.1$\pm$3.6 & -4.5$\pm$4.9 &  0.1$\pm$3.9 & -0.2$\pm$4.9 & -24.8$\pm$9.7 & ND\\
19 & 8.8$\pm$4.0 & 8.2$\pm$5.4 & 6.6$\pm$4.3 & 6.5$\pm$5.4 & 18.6$\pm$10.6 & -1.5$\pm$4.0  & 0.9$\pm$5.4 & -1.7$\pm$4.3 & -4.9$\pm$5.4 & 7.8$\pm$10.6 & DD\\ 
20 & -2.5$\pm$2.8 & 1.7$\pm$3.8 & -1.8$\pm$3.1 & -2.0$\pm$3.8 & -7.1$\pm$7.5 & 6.3$\pm$2.8 & 3.1$\pm$3.8 &  5.9$\pm$3.1 & 4.7$\pm$3.8 & 7.2$\pm$7.5 & DD\\
21 & -1.6$\pm$2.8 & -6.5$\pm$3.8 & -0.9$\pm$3.0 & 3.0$\pm$3.8 & -9.3$\pm$7.4 &  0.8$\pm$2.8 & 2.7$\pm$3.8 & -0.6$\pm$3.0  & 2.4$\pm$3.8  & 5.9$\pm$7.4 & DD\\
22 & -19.2$\pm$3.1 & -27.6$\pm$4.1 & -17.0$\pm$3.3 & -13.0$\pm$4.2 & -42.1$\pm$8.1 & -2.6$\pm$3.1 & -4.2$\pm$4.1 &  -2.6$\pm$3.3 & -0.9$\pm$4.1 & -2.3$\pm$8.1 & DD\\
23 & -18.9$\pm$3.1 & -24.5$\pm$4.1 & -16.2$\pm$3.3 & -11.6$\pm$4.1 & -30.8$\pm$8.1 & 3.8$\pm$3.1 & 3.2$\pm$4.1 &  4.1$\pm$3.3 & 1.6$\pm$4.1 & -0.1$\pm$8.1 & DD\\
24 & -11.2$\pm$2.7 & -18.0$\pm$3.7 & -11.8$\pm$3.0 & -6.5$\pm$3.7 & -9.5$\pm$7.2 & -1.0$\pm$2.7 & 3.3$\pm$3.7 &  -0.9$\pm$3.0 & -2.7$\pm$3.7 & -1.5$\pm$7.2 & DD\\
25 & -22.3$\pm$3.5 & -32.1$\pm$4.8 & -21.7$\pm$3.8 & -14.8$\pm$4.8 & -34.1$\pm$9.3 & 1.2$\pm$3.5 & 1.3$\pm$4.8 &  2.1$\pm$3.8 & 1.2$\pm$4.8 & -4.5$\pm$9.3  & DD\\
26 & -20.4$\pm$3.3 & -36.5$\pm$4.5 & -19.2$\pm$3.6 & -7.0$\pm$4.5 & -29.2$\pm$8.7 & 2.3$\pm$3.3 & 4.4$\pm$4.4  &  2.6$\pm$3.6 & 2.1$\pm$4.5 & 3.6$\pm$8.7  & DD \\
27 & -18.8$\pm$3.1 & -19.8$\pm$4.1 & -17.2$\pm$3.3 & -17.0$\pm$4.1 & -29.9$\pm$8.0 & -0.8$\pm$3.1 & -1.7$\pm$4.1 &  0.7$\pm$3.3 & -2.2$\pm$4.1 & -8.1$\pm$8.0 &  DD\\
28 & -18.7$\pm$3.1 & -27.1$\pm$4.2 & -18.1$\pm$3.4 & -8.4$\pm$4.2 & -26.7$\pm$8.2 & -3.4$\pm$3.1 & -1.8$\pm$4.2 &  -2.3$\pm$3.4 & -3.0$\pm$4.2 &  -10.4$\pm$8.2 & MD\\
29 & -17.4$\pm$4.2 & -24.8$\pm$5.7 & -18.3$\pm$4.6 & -13.5$\pm$5.7 & -13.6$\pm$11.2 & 2.3$\pm$4.2 &  2.4$\pm$5.7 &  1.8$\pm$4.6  & 4.5$\pm$5.7 & 3.7$\pm$11.2 & DD\\
30 & -11.2$\pm$3.9 & -14.0$\pm$5.3 & -9.3$\pm$4.2 & -12.3$\pm$5.3 & -35.0$\pm$10.4 & 0.9$\pm$3.9 & 5.3$\pm$5.3 &   1.8$\pm$4.2 & -0.3$\pm$5.3 & -16.0$\pm$10.4 & DD\\
31 & -19.5$\pm$3.9 & -24.2$\pm$5.2 & -18.0$\pm$4.2 & -8.1$\pm$5.2 & -11.8$\pm$10.2 & -1.2$\pm$3.9 &  -0.1$\pm$5.2 &  0.6$\pm$4.2 & -0.4$\pm$5.2 & -11.5$\pm$10.2 & DD\\
32 & -21.5$\pm$4.0 & -28.6$\pm$5.3 & -18.0$\pm$4.3 & -13.6$\pm$5.4 & -38.6$\pm$10.5 & 3.3$\pm$4.0 &  -0.7$\pm$5.3 &  3.9$\pm$4.3 & 5.5$\pm$5.4 & -4.6$\pm$10.5 &DD \\
33 & -20.6$\pm$3.8 & -22.3$\pm$5.1 & -20.1$\pm$4.1 & -16.4$\pm$5.1 & -37.6$\pm$9.9 & 6.5$\pm$3.8 &  13.9$\pm$5.1 & 5.0$\pm$4.1  &  -0.7$\pm$5.1 & 12.0$\pm$9.9  & DD\\
34 & -11.4$\pm$3.4 & -17.5$\pm$4.6 & -12.3$\pm$3.7 & -5.1$\pm$4.6 & -14.2$\pm$9.0 & -3.4$\pm$3.4 & -5.4$\pm$4.6 &  -4.9$\pm$3.7 &  -4.2$\pm$4.6  & 2.0$\pm$9.0 &DD \\
35 & -16.1$\pm$3.3 & -17.4$\pm$4.5 & -14.7$\pm$3.6 & -13.6$\pm$4.5 & -21.8$\pm$8.8 & -3.3$\pm$3.3 &  -8.1$\pm$4.5 &  -3.9$\pm$3.6 &  -3.2$\pm$4.5 & 3.0$\pm$8.8 & DD\\
36 & -16.7$\pm$3.4 & -21.1$\pm$4.6 & -16.1$\pm$3.7 & -12.2$\pm$4.6 & -19.9$\pm$9.0 & 2.9$\pm$3.4 &  4.4$\pm$4.6 &   2.3$\pm$3.7  &  4.5$\pm$4.6 &  16.8$\pm$9.0 & DD\\
37 & -13.5$\pm$3.3 & -18.6$\pm$4.5 & -12.0$\pm$3.6 & -7.0$\pm$4.5 & -19.6$\pm$8.7 & 0.6$\pm$3.3 &  2.8$\pm$4.5 &  -0.2$\pm$3.6 &  -2.9$\pm$4.5 & 6.9$\pm$8.7 & DD\\
38 & -4.9$\pm$2.3 & -7.9$\pm$3.1 & -4.2$\pm$2.5 & -1.3$\pm$3.1 & -6.2$\pm$6.1 & 0.7$\pm$2.3 &  3.3$\pm$3.1 &  0.6$\pm$2.5 &  -0.6$\pm$3.1 &  0.2$\pm$6.1 &  DD\\
39 & -7.3$\pm$2.2 & -11.4$\pm$2.9 & -5.1$\pm$2.4 & -6.2$\pm$2.9 & -21.1$\pm$5.7 & -0.9$\pm$2.2 &   -2.2$\pm$2.9  &  0.7$\pm$2.4 &  0.3$\pm$2.9  & -11.4$\pm$5.7 & DD\\
40 & -16.8$\pm$2.4 & -23.2$\pm$3.2 & -15.7$\pm$2.6 & -9.9$\pm$3.2 & -20.6$\pm$6.3 & 1.0$\pm$2.4 &   2.7$\pm$3.2 &  0.9$\pm$2.6 &  0.1$\pm$3.2 & -2.3$\pm$6.3 & DD\\
41 & -13.5$\pm$2.3 & -18.0$\pm$3.2 & -12.4$\pm$2.5 & -7.9$\pm$3.2 & -26.4$\pm$6.3 & -2.7$\pm$2.3 &  -4.2$\pm$3.2 &  -4.5$\pm$2.5 & -2.4$\pm$3.2  & 7.1$\pm$6.3 & DD\\
42 & -18.1$\pm$2.1 & -25.1$\pm$2.8 & -17.8$\pm$2.3 & -12.5$\pm$2.8 & -19.3$\pm$5.5 & -0.9$\pm$2.1 &  2.3$\pm$2.8 &  -1.8$\pm$2.3  &  -1.1$\pm$2.8 & 12.2$\pm$5.5  &  DD\\
43 & -8.8$\pm$2.1 & -11.4$\pm$2.8 & -7.6$\pm$2.2 & -9.3$\pm$2.8 & -16.9$\pm$5.4 & 2.8$\pm$2.1  &  1.4$\pm$2.8 &  3.5$\pm$2.2  &  2.9$\pm$2.8 & -1.1$\pm$5.4 & DD\\
44 & -3.7$\pm$2.0 & -5.1$\pm$2.7 & -4.0$\pm$2.2 & -2.6$\pm$2.7 & -3.2$\pm$5.3 &4.3$\pm$2.0  &  7.3$\pm$2.7  & 3.5$\pm$2.2  &  0.4$\pm$2.7 & 6.6$\pm$5.3  & DD\\
45 & -0.3$\pm$1.8 & -4.0$\pm$2.3 & -1.0$\pm$1.9 & -2.1$\pm$2.3 & -5.8$\pm$4.9 & 0.7$\pm$1.8 & -2.1$\pm$2.3 &  1.6$\pm$1.9  &  2.3$\pm$2.3 & -8.5$\pm$4.9  & DD\\
47 & 9.0$\pm$1.7 & 16.2$\pm$2.3 & 8.5$\pm$1.8 & 4.6$\pm$2.3 & 16.1$\pm$4.5 & -0.8$\pm$1.7  &  1.1$\pm$2.3 &  -1.6$\pm$1.8 &  -2.0$\pm$2.3  & 3.0$\pm$4.5 & DD\\
50 & 7.6$\pm$2.0 & 11.2$\pm$2.8 & 8.0$\pm$2.2 & 1.9$\pm$2.7 & 1.5$\pm$5.4 & -0.3$\pm$2.0  &  0.2$\pm$2.8 & 0.6$\pm$2.2  &  -0.7$\pm$2.7 & -6.1$\pm$5.4 & DD\\
51 & -13.3$\pm$1.7 & -18.7$\pm$2.4 & -12.9$\pm$1.9 & -8.7$\pm$2.3 & -17.0$\pm$4.5 & -0.3$\pm$1.7 &  1.5$\pm$2.3  &  0.1$\pm$1.9 &  0.9$\pm$2.3 &  -0.2$\pm$4.5 & DD\\
52 & -14.5$\pm$1.6 & -25.3$\pm$2.2 & -13.4$\pm$1.7 & -6.5$\pm$2.2 & -26.5$\pm$4.2 & -0.5$\pm$1.6 &  -1.2$\pm$2.2 & -1.1$\pm$1.7  &  -2.0$\pm$2.2 & -0.6$\pm$4.2 & DD\\
53 & -16.4$\pm$1.8 & -24.7$\pm$2.5 & -15.2$\pm$2.0 & -12.3$\pm$2.4 & -24.9$\pm$4.7 & -2.7$\pm$1.8  &  -5.0$\pm$2.4 & -2.6$\pm$1.9  &  -0.0$\pm$2.4 & -0.3$\pm$4.7 & DD\\
54 & -4.4$\pm$1.8 & -5.4$\pm$2.4 & -4.4$\pm$2.0 & -3.2$\pm$2.4 & -8.9$\pm$4.7 &  -1.2$\pm$1.8 &  -4.0$\pm$2.4 &   -0.8$\pm$2.0 &  1.9$\pm$2.4 & -6.2$\pm$4.7 & DD \\
55 & 6.9$\pm$1.8 & 11.1$\pm$2.5 & 7.1$\pm$2.0 & 5.6$\pm$2.5 & 8.2$\pm$4.9 &  -3.7$\pm$1.8 &  -5.1$\pm$2.5 & -3.5$\pm$2.0   &  -2.9$\pm$2.5 & -10.2$\pm$4.8 & ND\\
56 & 0.4$\pm$1.7 & -1.2$\pm$2.2 & 0.8$\pm$1.8 & 2.1$\pm$2.2 & -4.1$\pm$4.3 &  0.3$\pm$1.6 &  -2.1$\pm$2.2 &  2.1$\pm$1.8 &   2.9$\pm$2.2 &  -6.4$\pm$4.3  & DD\\
57 & -10.4$\pm$1.9 & -15.1$\pm$2.6 & -9.7$\pm$2.1 & -5.3$\pm$2.6 & -16.7$\pm$5.0 &  1.8$\pm$1.9 & 1.1$\pm$2.6  &  1.8$\pm$2.1 &  3.7$\pm$2.6 & 1.6$\pm$5.0 & DD\\
58 & -11.4$\pm$1.6 & -18.5$\pm$2.1 & -9.7$\pm$1.7 & -6.0$\pm$2.1 & -28.4$\pm$4.1 &  1.2$\pm$1.6 &  0.8$\pm$2.2 & 1.9$\pm$1.8  &  0.8$\pm$2.1 & -7.8$\pm$4.1 & DD\\
59 & 6.5$\pm$2.0 & 8.5$\pm$2.7 & 6.9$\pm$2.1 & 5.0$\pm$2.7 & 10.0$\pm$5.4 &  0.6$\pm$2.0 &  -4.0$\pm$2.7  & 0.6$\pm$2.1 &  3.7$\pm$2.7 & -4.6$\pm$5.4 & DD\\
60 & 6.4$\pm$1.7 & 9.6$\pm$2.3 & 4.4$\pm$1.8 & 3.1$\pm$2.3 & 15.9$\pm$4.6 &  -0.9$\pm$1.7  & -1.4$\pm$2.4 &   -1.3$\pm$1.8 &  -1.3$\pm$2.3 & 1.8$\pm$4.6 & DD \\
61 & -2.1$\pm$1.6 & -4.9$\pm$2.1 & -2.6$\pm$1.7 & -0.7$\pm$2.1 & 1.1$\pm$4.3 &   1.1$\pm$1.6 &  -0.4$\pm$2.2 &  0.7$\pm$1.7  &  -0.1$\pm$2.1 & -1.9$\pm$4.2 & DD\\
70 & -9.5$\pm$2.1 & -14.6$\pm$2.8 & -9.7$\pm$2.3 & -7.6$\pm$2.8 & -9.3$\pm$5.5 &   -1.1$\pm$2.1 & 0.9$\pm$2.8  &  -0.6$\pm$2.3 &  -2.5$\pm$2.8 & -5.0$\pm$5.5  & DD\\
71 & -7.5$\pm$1.8 & -17.0$\pm$2.5 & -7.0$\pm$2.0 & -3.1$\pm$2.5 & -12.8$\pm$4.8 & -0.4$\pm$1.8  &  -1.9$\pm$2.5 &  -0.7$\pm$2.0 &  -0.2$\pm$2.5 & -0.7$\pm$4.8 & DD\\
72 & -15.5$\pm$2.6 & -23.3$\pm$3.5 & -13.6$\pm$2.8 & -8.9$\pm$3.3 & -28.1$\pm$6.7 &  0.3$\pm$2.6 &  -1.0$\pm$3.5 & 0.6$\pm$2.8  &  1.4$\pm$3.2 & -2.1$\pm$6.7 & MD\\
73 & -18.5$\pm$2.0 & -23.7$\pm$2.7 & -16.9$\pm$2.1 & -9.9$\pm$2.7 & -27.1$\pm$5.3 &  -2.2$\pm$2.0 &  -2.0$\pm$2.7 & -2.1$\pm$2.1  &  1.6$\pm$2.7 & 0.4$\pm$5.3 & DD\\
74 & -10.7$\pm$1.9 & -13.5$\pm$2.6 & -10.3$\pm$2.1 & -6.8$\pm$2.6 & -12.0$\pm$5.1 &  1.1$\pm$1.9 &  2.2$\pm$2.6 & 1.5$\pm$2.1 & -2.2$\pm$2.6 & -1.2$\pm$5.1 & DD\\
75 & 4.7$\pm$1.8 & 6.3$\pm$2.4 & 4.3$\pm$1.9 & 2.5$\pm$2.4 & 10.1$\pm$4.6 &  -0.4$\pm$1.8 &    0.4$\pm$2.4 & 0.2$\pm$1.9 &  0.1$\pm$2.4 & -4.3$\pm$4.6 & DD\\
76 & 9.8$\pm$1.7 & 14.2$\pm$2.3 & 9.0$\pm$1.8 & 4.8$\pm$2.3 & 12.2$\pm$4.5 &  -1.8$\pm$1.7 &  0.3$\pm$2.4 & -2.7$\pm$1.8  &  -2.2$\pm$2.3  & 2.9$\pm$4.6 & DD\\
77 & 10.1$\pm$1.7 & 11.2$\pm$2.4 & 10.5$\pm$1.9 & 8.3$\pm$2.4 & 9.0$\pm$4.7 &   0.3$\pm$1.8 &   -1.6$\pm$2.4  &  1.2$\pm$1.9 & 0.2$\pm$2.4 & -11.6$\pm$4.8 & DD\\
78 & -17.0$\pm$1.8 & -25.8$\pm$2.5 & -16.1$\pm$2.0 & -9.8$\pm$2.5 & -22.0$\pm$4.9 &   0.3$\pm$1.8 &    -2.0$\pm$2.5 & 0.5$\pm$2.0 & 0.6$\pm$2.5 & -1.5$\pm$4.9 & DD\\
79 & -0.5$\pm$1.8 & -0.6$\pm$2.5 & -0.3$\pm$2.0 & -0.9$\pm$2.4 & -0.1$\pm$4.7 &  1.2$\pm$1.8 &  0.6$\pm$2.5 & 0.6$\pm$1.9 & -0.5$\pm$2.4 & 3.0$\pm$4.8 & DD\\
80 & 1.5$\pm$1.8 & 0.0$\pm$2.5 & 3.3$\pm$2.0 & 4.2$\pm$2.4 & -2.9$\pm$4.8 &   -1.1$\pm$1.8 &  0.2$\pm$2.5 & -1.9$\pm$2.0  & -1.6$\pm$2.4 &  4.3$\pm$4.8 & DD  \\
81 & 7.2$\pm$1.7 & 10.6$\pm$2.4 & 7.5$\pm$1.9 & 6.2$\pm$2.4 & 14.9$\pm$4.6 &  0.7$\pm$1.7 &  -0.7$\pm$2.4 & 0.0$\pm$1.9 & 0.7$\pm$2.4 & -3.5$\pm$4.6   & DD \\
\hline
82 & 0.7$\pm$1.7 & 1.3$\pm$2.5 & 0.3$\pm$1.8 & 1.4$\pm$2.2 & -1.8$\pm$5.3 & 1.2$\pm$1.7 &   2.7$\pm$2.5 & 1.5$\pm$1.8  & 0.6$\pm$2.2  &  -0.8$\pm$5.3 & DD\\
83 & -10.6$\pm$1.7 & -19.0$\pm$2.3 & -9.8$\pm$1.9 & -6.2$\pm$2.3 & -22.0$\pm$4.4 & -1.1$\pm$1.7  &  -1.6$\pm$2.3  &  -1.0$\pm$1.9 &  -1.9$\pm$2.3 & 2.6$\pm$4.4 & DD  \\
84 & -17.4$\pm$1.7 & -27.9$\pm$2.5 & -15.5$\pm$1.8 & -10.1$\pm$2.2 & -33.3$\pm$5.2 &  -0.6$\pm$1.7 & -1.8$\pm$2.5 & 0.0$\pm$1.8 & 1.1$\pm$2.2  & -5.5$\pm$5.2 & DD\\
85 & -19.5$\pm$1.8 & -26.5$\pm$2.6 & -19.2$\pm$1.9 & -14.5$\pm$2.3 & -24.8$\pm$5.6 &   -1.8$\pm$1.7 &  -1.0$\pm$2.6 & -2.6$\pm$1.9 & -2.4$\pm$2.3 & 8.9$\pm$5.5 & DD\\
87 & 4.7$\pm$1.8 & 5.1$\pm$2.4 & 3.6$\pm$1.9 & 1.5$\pm$2.4 & 6.7$\pm$4.7 &  3.1$\pm$1.7 &  0.0$\pm$2.4 & 2.8$\pm$1.9 &  2.8$\pm$2.4 & 5.2$\pm$4.7  & DD\\
88 & -17.0$\pm$2.6 & -19.5$\pm$3.5 & -16.0$\pm$2.8 & -13.6$\pm$3.5 & -19.7$\pm$6.8 &  3.2$\pm$2.6 &  0.0$\pm$3.5 & 3.9$\pm$2.8 &  3.5$\pm$3.5 & -2.3$\pm$6.9  & DD\\
89 & -6.7$\pm$2.3 & -9.7$\pm$3.2 & -7.1$\pm$2.5 & -4.8$\pm$3.1 & -4.7$\pm$6.2 &  -0.4$\pm$2.3 &  -0.3$\pm$3.2 &  0.2$\pm$2.5 & 0.1$\pm$3.1 & -5.5$\pm$6.2  & DD\\
90 & 3.4$\pm$2.2 & 8.9$\pm$3.1 & 2.5$\pm$2.4 & 0.6$\pm$2.7 & 6.2$\pm$5.9 &   1.0$\pm$2.2 &  2.4$\pm$3.1 & 0.7$\pm$2.3 & 1.4$\pm$2.8 & 3.2$\pm$5.9 & DD \\
91 & 3.9$\pm$2.4 & 8.8$\pm$3.4 & 3.1$\pm$2.6 & -0.1$\pm$3.3 & 6.9$\pm$6.6 & -1.2$\pm$2.4  &  1.8$\pm$3.4 &  -0.7$\pm$2.6 & -0.2$\pm$3.3 & -1.1$\pm$6.6 & DD\\
92 & -15.0$\pm$2.0 & -20.7$\pm$2.8 & -15.5$\pm$2.2 & -9.3$\pm$2.7 & -12.7$\pm$5.3 &  -0.4$\pm$2.0 &  1.6$\pm$2.7 & -2.1$\pm$2.2 & -1.2$\pm$2.7 &  6.5$\pm$5.3 & DD\\
93 & 6.2$\pm$1.7 & 11.6$\pm$2.3 & 6.3$\pm$1.8 & 5.2$\pm$2.3 & 8.9$\pm$4.4 &  0.7$\pm$1.7 &  3.1$\pm$2.3 & -0.1$\pm$1.8 & -1.3$\pm$2.3 & 6.5$\pm$4.4 & DD\\
94 & -1.6$\pm$1.6 & -1.4$\pm$2.2 & -0.3$\pm$1.7 & -1.0$\pm$2.1 & -4.8$\pm$4.2 &   -0.7$\pm$1.6 & -2.9$\pm$2.2 &  -0.5$\pm$1.7 & 1.0$\pm$2.1 & -2.4$\pm$4.1 & DD \\
95 & 0.3$\pm$1.6 & -0.4$\pm$2.2 & 0.2$\pm$1.8 & -1.4$\pm$2.1 & -4.6$\pm$4.2 &   3.9$\pm$1.6 &  4.0$\pm$2.2 & 3.1$\pm$1.8  & 1.9$\pm$2.1 & 7.9$\pm$4.2 & DD\\
96 & 9.7$\pm$1.8 & 14.9$\pm$2.4 & 9.1$\pm$1.9 & 7.0$\pm$2.3 & 17.0$\pm$4.7 &  1.4$\pm$1.8  &  4.6$\pm$2.4 & 2.1$\pm$1.9 & 0.8$\pm$2.3 & 0.2$\pm$4.7  & DD\\
\hline
\label{log_B}
\end{longtable}}

\end{document}